\documentclass[twocolumn,showpacs,preprintnumbers,amssymb,nofootinbib]{revtex4}

\usepackage{graphicx}
\usepackage{bm} 

\begin{document}


\title{Pair of accelerated black holes in a de Sitter
background: the dS C-metric}

\author{\'Oscar J. C. Dias}
\email{oscar@fisica.ist.utl.pt}
\author{Jos\'e P. S. Lemos}
\email{lemos@kelvin.ist.utl.pt}
\affiliation{
Centro Multidisciplinar de Astrof\'{\i}sica - CENTRA,
Departamento de F\'{\i}sica, Instituto Superior T\'ecnico,
Av. Rovisco Pais 1, 1049-001 Lisboa, Portugal
}%

\date{\today}

\begin{abstract}
Following the work of Kinnersley and Walker for flat spacetimes,
we have analyzed the anti-de Sitter C-metric in a previous paper.
In the de Sitter case, Podolsk\'y and Griffiths have established
that the de Sitter C-metric (dS C-metric) found by Pleba\'nski and
Demia\'nski describes a pair of accelerated black holes in the dS
background with the acceleration being provided (in addition to
the cosmological constant) by a strut that pushes away the two
black holes or, alternatively, by a string that pulls them. We
extend their analysis mainly in four directions. First, we draw
the Carter-Penrose diagrams of the massless uncharged dS C-metric,
of the massive uncharged dS C-metric and of the massive charged dS
C-metric. These diagrams allow us to clearly identify the presence
of two dS black holes and to conclude that they cannot interact
gravitationally. Second, we revisit the embedding of the dS
C-metric in the 5D Minkowski spacetime and we represent the motion
of the dS C-metric origin in the dS 4-hyperboloid as well as the
localization of the strut. Third, we comment on the physical
properties of the strut that connects the two black holes.
Finally, we find the range of parameters that correspond to
non-extreme black holes, extreme black holes, and naked particles.
\end{abstract}

\pacs{04.20.Jb,04.70.Bw,04.20.Gz}

\maketitle
\section{Introduction}

In a previous paper \cite{OscLem_AdS-C} we have analyzed in detail
the physical interpretation and properties of the anti-de Sitter
C-metric, i.e., the C-metric with a negative cosmological
constant, following the approach of Kinnersley and Walker
\cite{KW} for the flat C-metric. We have concluded that it
describes a pair of accelerated black holes when the acceleration
$A$ and the cosmological length $\ell$ are related by $A>1/\ell$.
The de Sitter C-metric (dS C-metric), i.e., the C-metric with a
positive cosmological constant, was introduced by Pleba\'nski and
Demia\'nski \cite{PlebDem}, and its physical interpretation was
first analyzed by Podolsk\'y and Griffiths \cite{PodGrif2}. A
special case of it has been applied in the study of the quantum
process of pair creation of black holes by Mann and Ross
\cite{MannRoss}, and by Booth and Mann \cite{BooMann} (see
\cite{Osc} for a review). The dS C-metric describes a pair of
accelerated black holes in the dS background with the acceleration
being provided (in addition to the cosmological constant) by a
strut that pushes away the two black holes or, alternatively, by a
string that pulls them. In this paper, we are interested in
further extend the physical interpretation of the de Sitter
C-metric mainly in four directions. First, we draw the
Carter-Penrose diagrams of the dS C-metric. These diagrams allow
us to clearly identify the presence of two dS black holes and to
conclude that they cannot interact gravitationally. Second, we
revisit the embedding of the dS C-metric in the 5D Minkowski
spacetime and we represent the motion of the dS C-metric origin in
the dS 4-hyperboloid as well as the localization of the strut.
Third, we discuss the physical properties of the strut that
connects the two black holes. Finally, we find the range of
parameters (cosmological constant, acceleration, mass and charge)
that correspond to non-extreme black holes, extreme black holes,
and naked particles.

The plan of this article is as follows. In section \ref{sec:Int}
we present the dS C-metric and analyze its curvature and conical
singularities. In section \ref{sec:PD} we study the causal
diagrams of the solution. In section \ref{sec:Phys_Interp} we give
and justify a physical interpretation to the solution. The
description of the solution in the dS 4-hyperboloid and the
physics of the strut are analyzed. Finally, in section
\ref{sec:Conc} concluding remarks are presented. For the
comparison with the C-metric in the anti-de Sitter background, as
well as for a fuller set of references on the C-metric, see
\cite{OscLem_AdS-C}.

\section{\label{sec:Int}GENERAL PROPERTIES}

In this section, we will briefly mention some general properties
of the dS C-metric that are well established. For details we ask
the reader to see, e.g., \cite{KW,OscLem_AdS-C}. The dS C-metric,
i.e., the C-metric with positive cosmological constant $\Lambda$,
has been obtained by Pleba\'nski and Demia\'nski \cite{PlebDem}.
For zero rotation and zero NUT parameter, the gravitational field
of the dS C-metric is given by (see \cite{OscLem_AdS-C})
\begin{equation}
 d s^2 = [A(x+y)]^{-2} (-{\cal F}dt^2+
 {\cal F}^{-1}dy^2+{\cal G}^{-1}dx^2+
 {\cal G}dz^2)\:,
 \label{C-metric}
 \end{equation}
 where
 \begin{eqnarray}
 & &{\cal F}(y) = -{\biggl (}\frac{1}{\ell^2A^2}+1{\biggl )}
                     +y^2-2mAy^3+q^2A^2y^4, \nonumber\\
 & &{\cal G}(x) = 1-x^2-2mAx^3-q^2 A^2 x^4\:,
 \label{FG}
 \end{eqnarray}
and the non-zero components of the electromagnetic vector
potential, $A_{\mu}dx^{\mu}$, are given by
\begin{eqnarray}
 A_{\rm t}=-e \,y \:, \;\;\;\;\;\;\;A_{\rm z}=g \, x  \:.
\label{potential}
\end{eqnarray}
 This solution depends on four parameters namely, the cosmological
 length $\ell^2\equiv3/\Lambda$, $A>0$ which
 is the acceleration of the black holes (see section
 \ref{sec:Phys_Interp}), and $m$ and $q$ which are
 interpreted  as the ADM mass and electromagnetic charge of the
 non-accelerated black holes, respectively (see Appendix). In general,
 $q^2=e^2+g^2$ with $e$ and $g$ being the electric and magnetic
 charges, respectively.

 The coordinates used in Eqs. (\ref{C-metric})-(\ref{potential}) hide
the physical interpretation of the solution. To understand the
physical properties of the solution we will introduce
progressively new coordinates more suitable to this propose,
following the approach of Kinnersley and Walker \cite{KW} and
Ashtekar and Dray \cite{AshtDray}. Although the alternative
 approach of Bonnor \cite{Bonnor1} simplifies in a way the interpretation, we
cannot use it were since the cosmological constant prevents such a
coordinate transformation into the Weyl form.

We start by defining a coordinate $r$ as
\begin{equation}
 r = [A(x+y)]^{-1} \:,
 \label{r}
 \end{equation}
which is interpreted as a radial coordinate. Indeed, calculating a
curvature invariant of the metric, namely the Kretschmann scalar,
\begin{eqnarray}
        R_{\mu\nu\alpha\beta}R^{\mu\nu\alpha\beta} &=&
        \frac{24}{\ell^2}
        +\frac{8}{r^8}{\biggl [}6m^2r^2+12m q^2(2Axr-1)r
                                 \nonumber \\
        & &
        +q^4(7-24Axr+24A^2x^2r^2){\biggr ]}
          \:,
                             \label{R2}
\end{eqnarray}
we conclude that it is equal to $24/\ell^2$ when the mass $m$ and
charge $q$ are both zero. Moreover, when at least one of these
parameters is not zero, the curvature invariant diverges at $r=0$,
revealing the presence of a curvature singularity. Finally, when
we take the limit $r\rightarrow \infty$, the curvature invariant
approaches the expected value for a spacetime which is
asymptotically dS.

We will consider only the values of $\Lambda$, $A$, $m$, and $q$
for which ${\cal G}(x)$ has at least two real roots,
$x_\mathrm{s}$ and $x_\mathrm{n}$ (say) and will demand that $x$
belongs to the range $[x_\mathrm{s},x_\mathrm{n}]$ where ${\cal
G}(x)\geq 0$. By doing this we guarantee that the metric has the
correct signature $(-+++)$ [see Eq. (\ref{C-metric})] and that the
angular surfaces $\Sigma$ ($t=$const and $r=$const) are compact
(see also the discussion in \cite{LetOliv}). In these angular
surfaces we now define two new coordinates,
\begin{eqnarray}
 \theta = \int_{x}^{x_\mathrm{n}}{\cal{G}}^{-1/2}dx \:,
\hspace{1cm} \phi &=& z/\kappa \:,
                             \label{ang}
\end{eqnarray}
where $\phi$ ranges between $[0,2\pi]$ and $\kappa$ is an
arbitrary positive constant which will be discussed soon. The
coordinate $\theta$ ranges between the north pole,
$\theta=\theta_\mathrm{n}=0$, and the south pole,
$\theta=\theta_\mathrm{s}$ (not necessarily at $\pi$). In general,
if we draw a small circle around the north or south pole, as the
radius goes to zero, the limit circunference/radius is not $2\pi$.
Indeed there is a deficit angle at the poles given by
$\delta_\mathrm{n/s}=2\pi [1-(\kappa/2)|d_x {\cal
G}|_{x_\mathrm{n/s}}]$. The value of $\kappa$ can be chosen in
order to avoid a conical singularity at one of the poles but we
cannot remove simultaneously the conical singularities at both
poles (for a more detailed analysis see \cite{OscLem_AdS-C}).

 Rewritten in terms of the new coordinates introduced in
Eq. (\ref{r}) and Eq. (\ref{ang}), the dS C-metric is given by
\begin{equation}
 d s^2 = r^2 [-{\cal F}(y)dt^2+
 {\cal F}^{-1}(y)dy^2+d\theta^2 + \kappa^2{\cal{G}}(x_{(\theta)})d\phi^2]\:,
 \label{dS C-metric}
 \end{equation}
where ${\cal F}(y)$ and ${\cal{G}}(x_{(\theta)})$ are given by Eq.
(\ref{FG}) and the time coordinate $t$ can take any value from the
interval $]-\infty,+\infty[$. When $m$ or $q$ are not zero there
is a curvature singularity at $r=0$. Therefore, we restrict the
radial coordinate to the range $[0,+\infty[$. On the other hand,
we  have restricted $x$ to belong to the range
$[x_\mathrm{s},x_\mathrm{n}]$ where ${\cal G}(x)\geq 0$. From
$Ar=(x+y)^{-1}$ we then conclude that $y$ must belong to the range
$-x\leq y < +\infty$. Indeed, $y=-x$ corresponds to $r=+\infty$,
and $y=+\infty$ to $r=0$. Note however that when both $m$ and $q$
vanish there are no restrictions on the ranges of $r$ and $y$
(i.e., $-\infty < r < +\infty$ and $-\infty < y < +\infty$) since
in this case there is no curvature singularity at the origin of
$r$ to justify the constraint on the coordinates.

\section{\label{sec:PD} Causal Structure}

In this section we analyze the causal structure of the solution.
The original dS C-metric, Eq. (\ref{dS C-metric}), is not
geodesically complete. To obtain the maximal analytic spacetime,
i.e., to draw the Carter-Penrose diagrams we will introduce the
usual null Kruskal coordinates. The description of the solution
depends crucially on the values of $m$ and $q$. We will consider
the three most relevant solutions, namely: {\it A. Massless
uncharged solution} ($m =0$, $q=0$), {\it B. Massive uncharged
solution} ($m>0$, $q=0$), and {\it C. Massive charged solution}
($m \geq0$, $q\neq0$).

\subsection{\label{sec:PD A.1}
 \textbf{Massless uncharged solution ($\bm{m=0, q=0}$)}}

\begin{figure}[t]
\includegraphics*[height=1.6in]{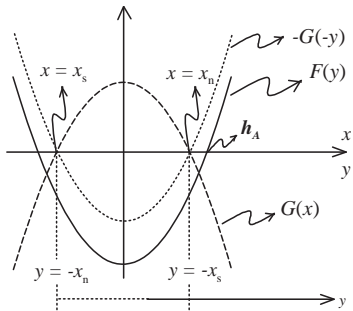}
\caption{\label{g1}
 Shape of ${\cal G}(x)$ and ${\cal F}(y)$ for the
 $m=0, q=0$ dS C-metric. The allowed range of $x$ is between
$x_\mathrm{s}=-1$ and $x_\mathrm{n}=+1$ where ${\cal G}(x)$ is
positive and compact. The range of $y$ is restricted to $-x\leq y
< +\infty$. The presence of an accelerated horizon is indicated by
$h_A$. It coincides with the cosmological horizon of the solution
and has a non-spherical shape.}
\end{figure}

In this case we have $x \in [x_\mathrm{s}=-1,x_\mathrm{n}=+1]$,
$x=\cos \theta$, ${\cal G}=1-x^2=\sin^2 \theta$, $\kappa=1$ and
\begin{eqnarray}
 {\cal F}(y) = y^2-y_+^2 \;\;\;\;\;\;\mathrm{with}\;\;\;\;\;\;
 y_+=\sqrt{1+\frac{1}{\ell^2A^2}} \:.
 \label{F1}
 \end{eqnarray}
The general behavior of these functions for this case is
represented in Fig. \ref{g1}.

 The angular surfaces $\Sigma$ ($t=$const and
$r=$const) are spheres and both the north and south poles are free
of conical singularities. The origin of the radial coordinate $r$
has no curvature singularity and therefore both $r$ and $y$ can
lie in the range $]-\infty,+\infty[$. However, in the realistic
general case, where $m$ or $q$ are non-zero, there is a curvature
singularity at $r=0$. Since the discussion of the present section
is only a preliminary to that of the massive general case we will
treat the origin $r=0$ as if it had a curvature singularity and
thus we admit that $r$ belongs to the range $[0,+\infty[$ and $y$
lies in the region $-x\leq y < +\infty$. We leave a discussion on
the extension to negative values of $r$ to section \ref{sec:PI}.

To construct the Carter-Penrose diagram we first introduce the
advanced and retarded Finkelstein-Eddington null coordinates,
\begin{eqnarray}
 u=t-y_* \:;    \;\;\;\;\;\;   v=t+y_* \:,
 \label{uv}
 \end{eqnarray}
where the tortoise coordinate coordinate is
\begin{eqnarray}
y_*=\int {\cal F}^{-1}dy=\frac{1}{2y_+} \ln{{\biggl
|}\frac{y-y_+}{y+y_+}{\biggl |}}=\frac{1}{2}(v-u) \:.
 \label{y*}
\end{eqnarray}
and both $u$ and $v$ belong to the range $]-\infty,+\infty[$. In
these coordinates the metric is given by
\begin{equation}
 d s^2 = r^2 [-{\cal F}dudv+
 d\theta^2 + \sin^2\!\theta \, d\phi^2]\:,
 \label{A.1.1}
 \end{equation}
 and his adjusted to the grid of null geodesics.
 Nevertheless, the metric still has the presence of a coordinate
 singularity at the root $y_+$ of ${\cal F}$. To overcome this
 undesirable situation we have to introduce the Kruskal
 coordinates.
The demand that $y$ must belong to the range $[-x;+\infty[$
implies that we have a region I, $-x\leq y < +y_+$, where ${\cal
F}(y)$ is negative and a region II, $y_+<y<+\infty$, where ${\cal
F}(y)$ is positive (see Fig. \ref{g1}). There is a Rindler-like
acceleration horizon ($r_A$) at $y=y_+$. This acceleration horizon
coincides with the cosmological horizon. In region I one sets the
Kruskal coordinates $u'=+e^{-\lambda u}$ and $v'=+e^{+\lambda v}$
so that $u'v'=+e^{2\lambda y_*}$. In region II one defines
$u'=-e^{-\lambda u}$ and $v'=+e^{+\lambda v}$ in order that
$u'v'=-e^{2\lambda y_*}$. We set $\lambda\equiv y_+$. In both
regions the product $u'v'$ is given by
\begin{eqnarray}
u'v'=-\frac{y-y_+}{y+y_+} \:,
 \label{u'v'}
\end{eqnarray}
and the metric (\ref{A.1.1}) expressed in terms of the Kruskal
coordinates is given by
\begin{eqnarray}
 d s^2 &=& r^2 {\biggl [}\frac{1}{y_+^2}\frac{{\cal F}}{u'v'}du'dv'+
 d\theta^2 + \sin^2\!\theta \,d\phi^2 {\biggr ]} \nonumber \\
      &=&  r^2 {\biggl [}-\frac{(y+y_+)^2}{y_+^2}du'dv'+
 d\theta^2 + \sin^2\!\theta \,d\phi^2 {\biggr ]} \:, \nonumber \\
 \label{A.1.2}
 \end{eqnarray}
with $y$ and $Ar=(x+y)^{-1}$ regarded as functions of $u'$ and
$v'$,
\begin{eqnarray}
 y=y_+\frac{1-u'v'}{1+u'v'}\:, \;\;\;\;
 Ar=\frac{1+u'v'}{(y_+ +x)-u'v'(y_+ -x)} \:.
 \label{y,r}
 \end{eqnarray}
The Kruskal coordinates in both regions were chosen in order to
obtain a negative value for the factor ${\cal F}/(u'v')$, which
appears in the metric coefficient $g_{u'v'}$. The value of
constant $\lambda$ was selected in order that the limit of ${\cal
F}/(u'v')$ as $y \to y_+$ stays finite and different from zero. By
doing this, we have removed the coordinate singularity that was
present at the root $y_+$ of ${\cal F}$ [see Eq. (\ref{A.1.1})].
So, the metric is now well-behaved in the whole range $-x\leq y
<+\infty$ or $0\leq r<+\infty$. At the edges of the interval
allowed for $r$, the product $u'v'$ takes the values
\begin{eqnarray}
 \lim_{r \to 0} u'v'=-1\:, \;\;\;\;\;
 \lim_{r \to +\infty} u'v'=\frac{y_+ + x}{y_+ - x}>0 \;
 \mathrm{and \; finite} \:. \nonumber \\
 \label{lim u'v'}
 \end{eqnarray}
So, the original massless uncharged dS C-metric is described by
the spacetime (\ref{A.1.2}) subjected to the following coordinates
ranges,
\begin{eqnarray}
 \hspace{-0.5cm} & & \hspace{-0.5cm}
 0 \leq \phi < 2\pi\:, \;\;\; -1 \leq x \leq +1 \:,\;\;\; u'<0\:,
 \;\;\; v'>0 \:,\;\;\;          \\
 \hspace{-0.5cm} & & \hspace{-0.5cm}
  -1\leq u'v'<\frac{y_+ + x}{y_+ - x}  \:.
 \label{ranges u'v'}
 \end{eqnarray}
 This spacetime is however geodesically incomplete. To obtain the
 maximal analytical extension one allows the Kruskal coordinates
 to take also the values $u'\geq 0$ and $v'\leq 0$ as soon as the
 condition (\ref{ranges u'v'}) is satisfied.

\begin{figure}  [t]
\includegraphics{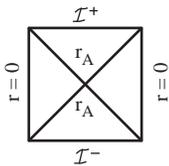}
   \caption{\label{Fig-1}
Carter-Penrose diagram concerning the $m=0, q=0$ dS C-metric
studied in section \ref{sec:PD A.1}. The accelerated horizon is
represented by $r_A$. It coincides with the cosmological horizon
and has a non-spherical shape. ${\cal I}^-$ and ${\cal I}^+$
represent respectively the past and future infinity ($r=+\infty$).
$r=0$ corresponds to $y=+\infty$ and $r=+\infty$ corresponds to
$y=-x$.
 }
\end{figure}

Finally, to construct the Carter-Penrose diagram one has to define
the Carter-Penrose coordinates by the usual arc-tangent functions
of $u'$ and $v'$: ${\cal{U}}=\arctan u'$ and ${\cal{V}}=\arctan
v'$, that bring the points at infinity into a finite position. The
Carter-Penrose diagram of the massless uncharged dS C-metric is
sketched in Fig. \ref{Fig-1}. $r=0$ is represented by a timelike
line while $r=+\infty$ is a spacelike line (with ${\cal I}^-$ and
${\cal I}^+$ representing, respectively, the past and future
infinity). The two mutual perpendicular straight null lines at
$45^{\rm o}$, $u'v'=0$, represent a Rindler-like accelerated
horizon.

To end this subsection it is important to remark that, contrary to
what happens in the C-metric with  $\Lambda<0$ \cite{OscLem_AdS-C}
and with $\Lambda=0$ \cite{KW}, the presence of the acceleration
in the $\Lambda>0$ C-metric does not introduce an extra horizon
relatively to the $A=0$ solution. Indeed, in the dS C-metric the
acceleration horizon coincides with the cosmological horizon that
is already present in the $A=0$ solution. However, in the $A=0$
solution the cosmological horizon has the topology of a round
sphere, while in the dS C-metric ($A\neq 0$) the presence of the
acceleration induces a non-spherical shape in the acceleration
(cosmological) horizon. This conclusion is set from the expression
of the radius of the horizon, $r_A=A^{-1}(x+y_+)^{-1}$. It varies
with the angular direction $x=\cos \theta$ and depends on the
value of $A$ [see Eq. (\ref{F1})]. Another important difference
between the causal structure of the dS C-metric and the causal
structure of the $\Lambda = 0$ and the $\Lambda<0$ cases is the
fact that the general features of the Carter-Penrose diagram of
the dS C-metric are independent of the angular coordinate $x=\cos
\theta$. Indeed, in the  $\Lambda<0$ case \cite{OscLem_AdS-C} and
in the $\Lambda=0$ case \cite{KW}, the Carter-Penrose diagram at
the north pole direction is substantially different from the one
along the south pole direction and different from the diagram
along the equator direction (see \cite{KW,OscLem_AdS-C}).

\subsection{\label{sec:PD A.2}
 \textbf{Massive uncharged solution ($\bm{m >0}$, $\bm{q=0}$)}}
\begin{figure} [t]
\includegraphics*[height=1.6in]{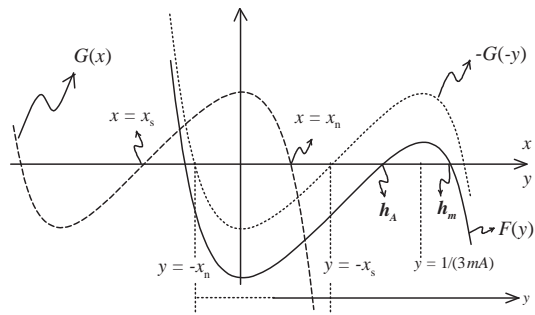}
\caption{\label{g2}
 Shape of ${\cal G}(x)$ and ${\cal F}(y)$ for the
 $27m^2A^2<1-9m^2\Lambda$, and $q=0$ dS C-metric (case (i) in the text).
The allowed range of $x$ is between $x_\mathrm{s}$ and
$x_\mathrm{n}$ where ${\cal G}(x)$ is positive and compact. The
range of $y$ is restricted to $-x\leq y < +\infty$. The presence
of an accelerated horizon (which coincides with the cosmological
horizon and has a non-spherical shape) is indicated by $h_A$ and
the Schwarzschild-like horizon by $h_m$. For completeness we
comment on two other cases studied in the text: for
$27m^2A^2=1-9m^2\Lambda$ (case (ii) in the text), ${\cal F}(y)$ is
zero at its local maximum, i.e., $h_A$ and $h_m$ coincide. For
$27m^2A^2>1-9m^2\Lambda$ (case (iii) in the text), ${\cal F}(y)$
is always negative in the allowed range of $y$.
 }
\end{figure}
The construction of the Carter-Penrose diagram for the $m> 0$ dS
C-metric follows up directly from the last subsection. We will
consider the small mass or acceleration case, i.e., we require
$27m^2A^2<1$ and we also demand $x$ to belong to the range
$[x_\mathrm{n},x_\mathrm{s}]$ (represented in Fig. \ref{g2} and
such that $x_\mathrm{s} \to -1$ and $x_\mathrm{n} \to +1$ when $mA
\to 0$) where ${\cal G}(x)\geq 0$. By satisfying the two above
conditions we endow the $t=$const and $r=$const surfaces $\Sigma$
with the topology of a compact surface. For $27m^2A^2 \geq 1$ this
surface is an open one and will not be discussed (see however
\cite{LetOliv}).

\begin{figure}[b]
\includegraphics*[height=2cm]{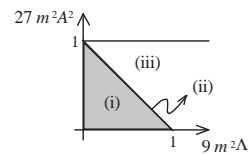}
   \caption{\label{Fig-RangeM}
Allowed ranges of the parameters $\Lambda, A$, and $m$ for the
cases (i), (ii), and (iii) of the uncharged massive dS C-metric
discussed in the text of section \ref{sec:PD A.2}.
 }
\end{figure}

Now we turn our attention to the behavior of function ${\cal
F}(y)$. We have to consider three distinct cases (see Fig.
\ref{Fig-RangeM}), namely: (i) pair of non-extreme
dS-Schwarzschild black holes ($27m^2A^2<1-9m^2\Lambda$) for which
${\cal F}(y=1/3mA)>0$ (see Fig. \ref{g2}), (ii) pair of extreme
dS-Schwarzschild black holes ($27m^2A^2=1-9m^2\Lambda$) for which
${\cal F}(y=1/3mA)=0$, and (iii) case $27m^2A^2>1-9m^2\Lambda$ for
which ${\cal F}(y)$ is always negative in the allowed range for
$y$. This last case represents a naked particle and will not be
discussed further. Notice that when we set $A=0$ in the above
relations we get the known results \cite{Lake} for the
non-accelerated dS spacetime, namely: for $9m^2\Lambda<1$ we have
the non-extreme dS-Schwarzschild solution and for $9m^2\Lambda=1$
we get the extreme dS-Schwarzschild solution. In what follows we
will draw the Carter-Penrose diagrams of cases (i) and (ii).

\begin{figure}[t]
\includegraphics*[height=4.8cm]{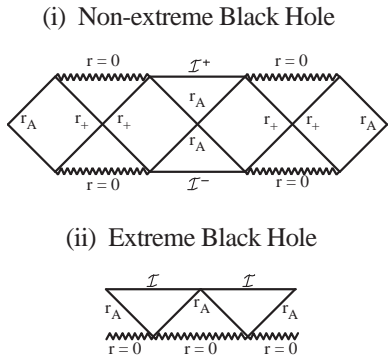}
   \caption{\label{Fig-2}
(i) Carter-Penrose diagram of the $27m^2A^2<1-9m^2\Lambda$, and
$q=0$ dS C-metric discussed in case (i) of section \ref{sec:PD
A.2}. The zigzag line represents a curvature singularity, the
accelerated horizon is represented by $r_A$. It coincides with the
cosmological horizon and has a non-spherical shape. The
Schwarzschild-like horizon is sketched as $r_+$. $r=0$ corresponds
to $y=+\infty$ and $r=+\infty$ (${\cal I}^-$ and ${\cal I}^+$)
corresponds to $y=-x$. (ii) Carter-Penrose diagram of the
degenerate case (ii), $27m^2A^2=1-9m^2\Lambda$ and $q=0$,
discussed in the text of section \ref{sec:PD A.2}. The accelerated
horizon $r_A$ coincides with the Schwarzschild-like horizon $r_+$.
 }
\end{figure}

\vspace{0.1 cm} (i) {\it Pair of non-extreme dS-Schwarzschild
black holes} ($27m^2A^2<1-9m^2\Lambda$): the technical procedure
to obtain the Carter-Penrose diagram is similar to the one
described along section \ref{sec:PD A.1}. In what concerns the
physical conclusions, we will see that the essential difference is
the presence of an extra horizon, a Schwarzschild-like horizon
($r_+$) due to the non-vanishing mass parameter, in addition to
the accelerated Rindler-like horizon ($r_A$). Another important
difference, as stated in section \ref{sec:Int}, is the presence of
a curvature singularity at the origin of the radial coordinate and
the existence of a conical singularity at one of the poles. The
Carter-Penrose diagram is drawn in Fig. \ref{Fig-2}.(i) and has a
structure that can be divided into left, middle and right regions.
The middle region contains the spacelike infinity (with ${\cal
I}^-$ and ${\cal I}^+$ representing, respectively, the past and
future infinity) and an accelerated Rindler-like horizon,
$r_A=[A(x-x_-)]^{-1}$, that was already present in the $m=0$
corresponding diagram [see Fig. \ref{Fig-1}]. The left and right
regions both contain a spacelike curvature singularity at $r=0$
and a Schwarzschild-like horizon, $r_+$. This diagram is analogous
to the one of the non-accelerated ($A=0$) dS-Schwarzschild
solution. However, in the $A=0$ solution the cosmological and
black hole horizons have the topology of a round sphere, while in
the dS C-metric ($A\neq 0$) the presence of the acceleration
induces a non-spherical shape in the acceleration horizon (that
coincides with the cosmological horizon) and in the black hole
horizon. Indeed, notice that once we find the zero, $y_h$, of
${\cal F}(y)$ that corresponds to an accelerated or black hole
horizon, the position of these horizons depends on the angular
coordinate $x$ since $r_h=[A(x+y_h)]^{-1}$. In section
\ref{sec:PI-mass} we will justify that this solution describes a
pair of accelerated dS-Schwarzschild black holes.

\vspace{0.1 cm} (ii) {\it Pair of extreme dS-Schwarzschild black
holes} ($27m^2A^2=1-9m^2\Lambda$): for this range of values, we
have a degenerate case in which the size of the black hole horizon
approaches and equals the size of the acceleration horizon. In
this case, as we shall see in section \ref{sec:PI-mass}, the dS
C-metric describes a pair of accelerated extreme dS-Schwarzschild
black holes. The Carter-Penrose diagram of this solution is
sketched in Fig. \ref{Fig-2}.(ii). It should be noted that for
this sector of the solution, and as occurs with the $A=0$ case,
there is an appropriate limiting procedure \cite{OscLemNariai}
that takes this solution into the Nariai C-metric, i.e., the
accelerated counterpart of the Nariai solution \cite{Nariai}.

\subsection{\label{sec:PD A.3}
\textbf{Massive charged solution ($\bm{m >0}$, $\bm{q\neq0}$)}}
\begin{figure} [b]
\includegraphics*[height=1.6in]{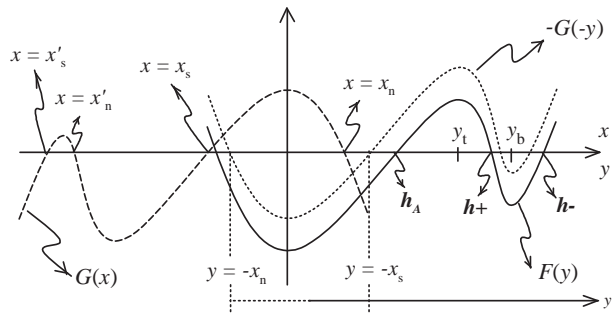}
\caption{\label{g3}
 Shape of ${\cal G}(x)$ and ${\cal F}(y)$ for the non-extreme
charged massive dS C-metric (case (i) in the text of section
\ref{sec:PD A.3}). The allowed range of $x$ is between
$x_\mathrm{s}$ and $x_\mathrm{n}$ where ${\cal G}(x)$ is positive
and compact.  The presence of an accelerated horizon is indicated
by $h_A$ and the inner and outer charged horizons by $h-$ and
$h+$. In the extreme cases, $h-$ and $h+$ [case (ii)] or $h-_A$
and $h+$ [case (iii)] superpose each other and in the naked case
[case (iv) and (v)] ${\cal F}(y)$ has only one zero in the allowed
range of $y$.
 }
\end{figure}

\begin{figure}[t]
\includegraphics*[height=9cm]{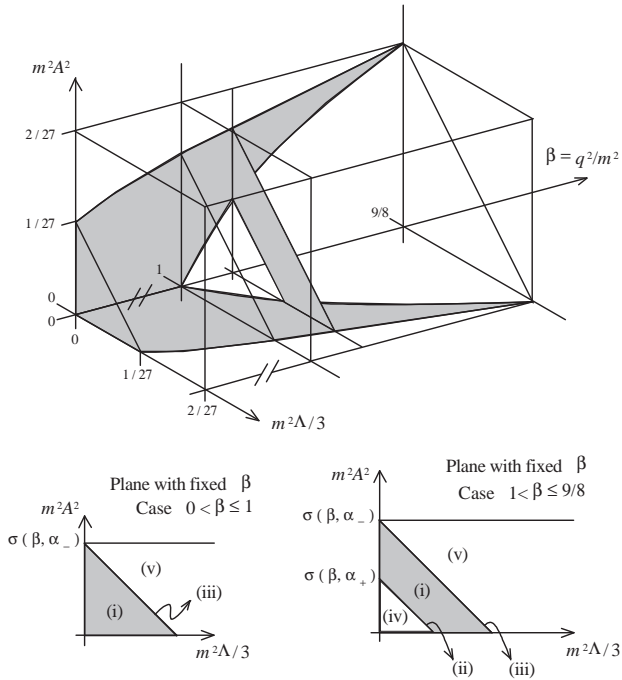}
   \caption{\label{Fig-RangeQ}
Allowed ranges of the parameters $\Lambda, A, m, \beta\equiv
q^2/m^2$ for the cases (i), (ii), (iii), (iv), and (v) of the
charged massive dS C-metric discussed in the text of section
\ref{sec:PD A.3}.
 }
\end{figure}

When both the mass and charge parameters are non-zero, depending
on the values of the parameters, ${\cal G}(x)$ can be positive in
a single compact interval, $]x_\mathrm{s},x_\mathrm{n}[$, or in
two distinct compact intervals, $]x_\mathrm{s},x_\mathrm{n}[$ and
$]x'_\mathrm{s},x'_\mathrm{n}[$, say. We require that $x$ belongs
to the interval $[x_\mathrm{s},x_\mathrm{n}]$ (sketched in Fig.
\ref{g3}) for which the charged solutions are in the same sector
of those we have analyzed in the last two subsections when $q \to
0$. Defining
\begin{eqnarray}
& & \beta \equiv \frac{q^2}{m^2}\:,  \;\;\; 0<\beta\leq
\frac{9}{8} \:, \;\;\;\;\;\alpha_{\pm} \equiv
1 \pm \sqrt{1-\frac{8}{9}\beta} \:, \nonumber \\
 & &
\sigma(\beta,\alpha_{\pm})=
\frac{(4\beta)^2(3\alpha_{\pm})^2-8\beta(3\alpha_{\pm})^3+\beta(3\alpha_{\pm})^4}
{(4\beta)^4} \:,  \nonumber \\
 \label{beta}
\end{eqnarray}
the above requirement is fulfilled by the parameter range
$m^2A^2<\sigma(\beta,\alpha_-)$. Now we look into the behavior of
the function ${\cal F}(y)$. Depending on the sign of ${\cal F}(y)$
at $y_{\rm t}$ and $y_{\rm b}$  (with $y_{\rm
t}=\frac{3\alpha_-}{4\beta mA}$ and
 $y_{\rm b}=\frac{3\alpha_+}{4\beta mA}$ being the points represented in
Fig. \ref{g3} where the derivative of ${\cal F}(y)$ vanishes) we
can group the solutions into five different relevant physical
classes, namely: (i) ${\cal F}(y_{\rm t})>0$ and ${\cal F}(y_{\rm
b})<0$, (ii) ${\cal F}(y_{\rm t})>0$ and ${\cal F}(y_{\rm b})=0$,
(iii) ${\cal F}(y_{\rm t})=0$ and ${\cal F}(y_{\rm b})<0$, (iv)
${\cal F}(y_{\rm t})>0$ and ${\cal F}(y_{\rm b})>0$, and (v)
${\cal F}(y_{\rm t})<0$ and ${\cal F}(y_{\rm b})<0$. The ranges of
parameters $\Lambda, A, m$, and $\beta$ that correspond to these
five cases are identified in Fig. \ref{Fig-RangeQ}.
Condition ${\cal F}(y_{\rm t})\geq 0$ requires
 $m^2A^2 \leq \sigma(\beta,\alpha_-)-m^2\Lambda/3$ and
 ${\cal F}(y_{\rm b})\leq 0$ is satisfied by
 $m^2A^2 \geq \sigma(\beta,\alpha_+)-m^2\Lambda/3$.
We have $\sigma(\beta,\alpha_-)>\sigma(\beta,\alpha_+)$ except at
$\beta=9/8$ where these two functions are equal;
$\sigma(\beta,\alpha_-)$ is always positive; and
$\sigma(\beta,\alpha_+)<0$ for $0<\beta<1$ and
$\sigma(\beta,\alpha_+)>0$ for $1<\beta\leq 9/8$. Case (i) has
three horizons, the acceleration horizon $h_A$ and the inner
($h_-$) and outer ($h_+$) charged horizons and is the one that is
exactly represented in Fig. \ref{g3} ($h_A\neq h_+ \neq h_-$); in
case (ii) the inner horizon and outer horizon coincide ($h_+
\equiv h_-$) and are located at $y_{\rm b}$ ($h_A$ is also
present); in case (iii) the acceleration horizon and outer horizon
coincide ($h_A \equiv h_+$) and are located at $y_{\rm t}$ ($h_-$
is also present); finally in cases (iv) and (v) there is a single
horizon  $h_A \equiv h_+ \equiv h_-$.
  As will be seen, case (i) describes a pair of accelerated
dS$-$Reissner-Nordstr\"{o}m (dS-RN) black holes, case (ii)
describes a pair of extreme dS-RN black holes in which the inner
and outer charged horizons become degenerated, case (iii)
describes a pair of extreme dS-RN black holes in which
acceleration horizon and outer charged horizon  become
degenerated, and cases (iv) and (v) describe a pair of naked
charged particles. It should be noted that for the sector (iii) of
the solution, and as occurs with the $A=0$ case,  there is an
appropriate limiting procedure \cite{OscLemNariai} that takes this
solution into the charged Nariai C-metric, i.e., the accelerated
counterpart of the charged Nariai solution \cite{MannRoss,Nariai}.

\begin{figure} [b]
\includegraphics*[height=14cm]{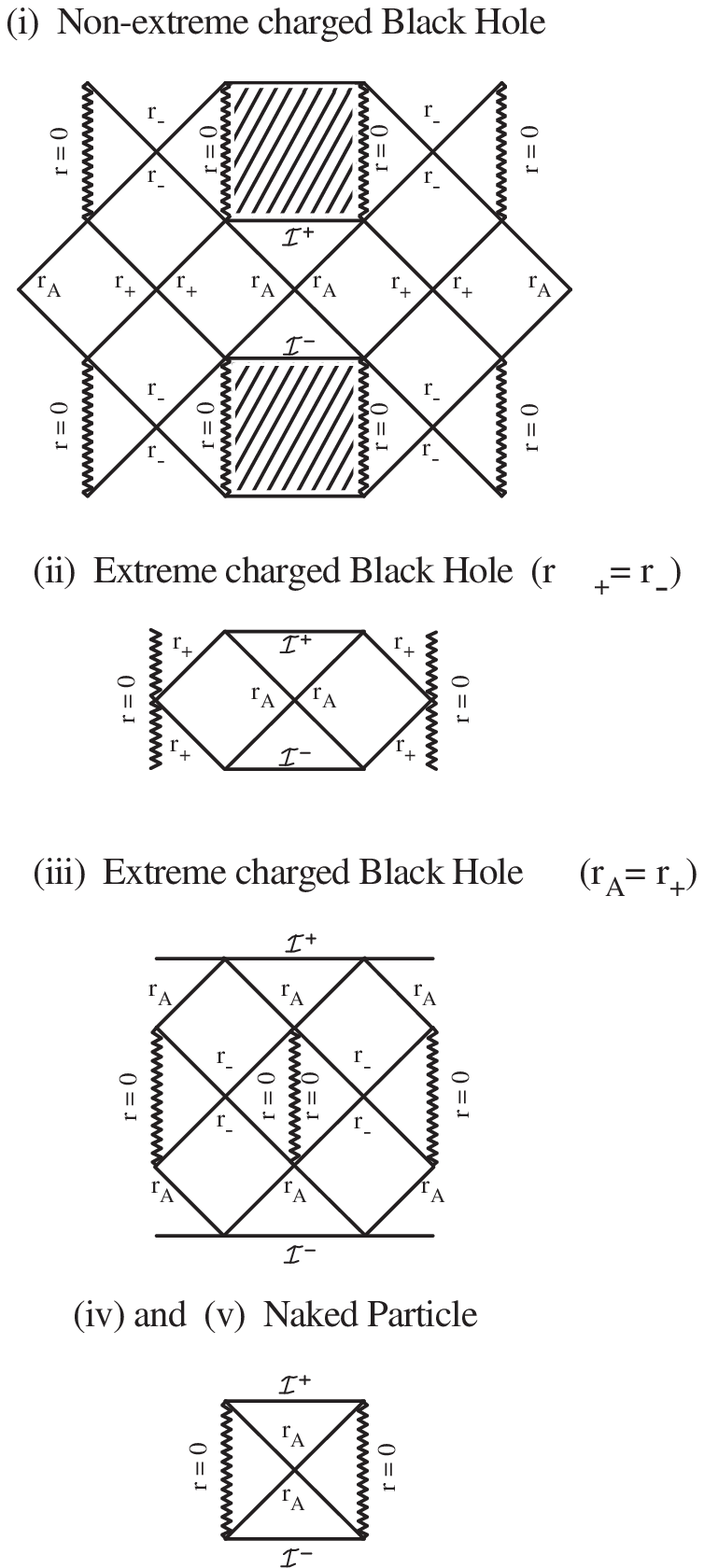}
   \caption{\label{Fig-3}
Carter-Penrose diagrams of cases (i), (ii), (iii), and (iv) and
(v) of the charged massive dS C-metric. The zigzag line represents
a curvature singularity, an accelerated horizon is represented by
$r_A$, the inner and outer charge associated horizons are sketched
as $r_-$ and $r_+$. ${\cal I}^-$ and ${\cal I}^+$ represent
respectively the past and future infinity ($r=+\infty$). $r=0$
corresponds to $y=+\infty$ and $r=+\infty$ corresponds to $y=-x$.
 }
\end{figure}

The essential differences between the Carter-Penrose diagram of
the massive charged solutions and the diagram of the massive
uncharged solutions are: (1) the curvature singularity is now
represented by a timelike line rather than a spacelike line, (2)
excluding the extreme and naked cases, there are now (in addition
to the accelerated Rindler-like horizon, $r_A$) not one but two
extra horizons, the expected inner ($r_-$) and outer ($r_+$)
horizons associated to the charged character of the solution.

The Carter-Penrose diagram of case (i) is drawn in Fig.
\ref{Fig-3}.(i) and has a structure that, as occurs in the massive
uncharged case, can be divided into left, middle and right
regions. The middle region contains the spacelike infinity and an
accelerated Rindler-like horizon, $r_A$, that was already present
in the $q=0=m$ corresponding diagram (see Fig. \ref{Fig-1}). The
left and right regions both contain a timelike curvature
singularity ($r=0$), and an inner ($r_-$) and an outer ($r_+$)
horizons associated to the charged character of the solution. This
diagram is analogous the the one of the non-accelerated ($A=0$)
dS$-$Reissner-Nordstr\"{o}m solution. However, in the $A=0$
solution the cosmological and black hole horizons have the
topology of a round sphere, while in the dS C-metric ($A\neq 0$)
the presence of the acceleration induces a non-spherical shape in
the accelerated horizon (that coincides with the cosmological
horizon) and in the black hole horizons. Indeed, notice that once
we find the zero, $y_h$, of ${\cal F}(y)$ that corresponds to an
accelerated or black hole horizon, the position of these horizons
depends on the angular coordinate $x$ since $r_h=[A(x+y_h)]^{-1}$.
In Fig. \ref{Fig-3} are also represented the other cases (ii)-(v).
Again the accelerated horizon is in between two (extreme) black
holes in cases (ii) and (iii) and in between two naked particles
in cases (iv) and (v).

\section{\label{sec:Phys_Interp} PHYSICAL INTERPRETATION}

The parameter $A$ that is found in the dS C-metric is interpreted
as being an acceleration and the dS C-metric describes a pair of
black holes accelerating away from each other in a dS background.
In this section we will justify this statement.

In the Appendix it is shown that, when $A=0$, the general dS
C-metric, Eq. (\ref{dS C-metric}), reduces to the dS
($m=0\,,\,q=0$), to the dS-Schwarzschild ($m>0\,,\,q=0$), and to
the dS$-$Reissner-Nordstr\"{o}m solutions ($m=0\,,\,q\neq0$).
Therefore, the parameters $m$ and $q$ are, respectively, the ADM
mass and ADM electromagnetic charge of the non-accelerated black
holes. Moreover, if we set the mass and charge parameters equal to
zero, even when $A\neq 0$, the Kretschmann scalar
 [see Eq. (\ref{R2})] reduces to the value expected for the dS spacetime.
This indicates that the massless uncharged dS C-metric is a dS
spacetime in disguise.

In this section, we will first interpret case {\it A. Massless
uncharged solution} ($m =0$, $q=0$), which is the simplest, and
then with the acquired knowledge we interpret cases {\it B.
Massive uncharged solution} ($m>0$, $q=0$) and {\it C. Massive
charged solution} ($m>0$, $q\neq0$). We will interpret the
solution following two complementary descriptions, the four
dimensional (4D) one and the five dimensional (5D).

\subsection{\label{sec:PI} \textbf{Description of
the $\bm{m =0}$, $\bm{q = 0}$ solution}}
\begin{figure} [ht]
\includegraphics*[height=2.3cm]{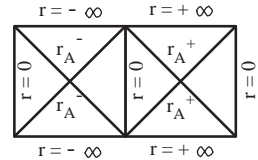}
   \caption{\label{Fig-dS}
Extending the Carter-Penrose diagram of Fig. \ref{Fig-1} to
negative values of $r$, we obtain the dS spacetime with its origin
being accelerated. $r_A^- =[A(x-y_+)]^{-1}<0$ and
$r_A^+=[A(x+y_+)]^{-1}>0$.
 }
\end{figure}

{\it The 4-Dimensional description}:

As we said in \ref{sec:PD A.1}, when $m=0$ and $q=0$ the origin of
the radial coordinate $r$ defined in Eq. (\ref{r}) has no
curvature singularity and therefore $r$ has the range
$]-\infty,+\infty[$. However, in the realistic general case, where
$m$ or $q$ are non-zero, there is a curvature singularity at $r=0$
and since the discussion of the massless uncharged solution was
only a preliminary to that of the massive general case, following
\cite{AshtDray}, we have treated the origin $r=0$ as if it had a
curvature singularity and thus we admitted that $r$ belongs to the
range $[0,+\infty[$. In these conditions we obtained the causal
diagram of Fig. \ref{Fig-1}. Note however that one can make a
further extension to include the negative values of $r$, enlarging
in this way the range accessible to the Kruskal coordinates $u'$
and $v'$. By doing this procedure we obtain the causal diagram of
the dS spacetime, represented in Fig. \ref{Fig-dS}.

Now, we want to clearly identify the parameter $A$ that appears in
the dS C-metric with the acceleration of its origin. To achieve
this aim, we recover the massless uncharged dS C-metric defined by
Eq. (\ref{C-metric}) and Eq. (\ref{FG}) (with $m=0$ and $q=0$),
and after performing the following coordinate transformation
\cite{PodGrif2}
\begin{eqnarray}
& & \tau=\frac{\sqrt{1+\ell^2A^2}}{A} t \:,  \;\;\;\;\;
    \rho=\frac{\sqrt{1+\ell^2A^2}}{A} \frac{1}{y} \:, \nonumber \\
& & \theta = \arccos{x} \:,
     \;\;\;\;\; \phi = z \:,
  \label{transf-int}
  \end{eqnarray}
we can rewrite the massless uncharged dS C-metric as
\begin{eqnarray}
 d s^2 = \frac{1}{\gamma^2}
 {\biggl [}-(1-\rho^2/\ell^2)d\tau^2+
 \frac{d\rho^2}{1-\rho^2/\ell^2} +\rho^2 d\Omega^2 {\biggl ]},
\label{metric-int}
 \end{eqnarray}
 with $d\Omega^2=d\theta^2+\sin^2\theta d \phi^2$ and
 \begin{eqnarray}
 \gamma=\sqrt{1+\ell^2A^2} + A\rho \cos\theta \:.
 \label{gamma}
 \end{eqnarray}
At this point some remarks are convenient. The origin of the
radial coordinate $\rho$ corresponds to $y=+\infty$ and therefore
to $r=0$, where $r$ has been introduced in Eq. (\ref{r}). So, when
we consider the massive dS C-metric there will be a curvature
singularity at $\rho=0$ (see section \ref{sec:Int}). Moreover,
when we set $A=0$, Eq. (\ref{metric-int}) reduces to the usual dS
spacetime written in static coordinates.

To discover the meaning of the parameter $A$ we consider the 4D
timelike worldlines described by an observer with $\rho=$const,
$\theta=0$ and $\phi=0$ (see \cite{PodGrif2}). These are given by
$x^{\mu}(\lambda)=(\gamma \ell
\lambda/\sqrt{\ell^2-\rho^2},\rho,0,0)$, were $\lambda$ is the
proper time of the observer and the 4-velocity
$u^{\mu}=dx^{\mu}/d\lambda$ satisfies $u_{\mu}u^{\mu}=-1$. The
4-acceleration of these observers,
$a^{\mu}=(\nabla_{\nu}u^{\mu})u^{\nu}$, has a magnitude given by
\begin{eqnarray}
 |a_4|=\sqrt{a_{\mu}a^{\mu}}=\frac{\rho\sqrt{1+\ell^2A^2}+\ell^2A}
 {\ell\sqrt{\ell^2-\rho^2}}\:.
 \label{a}
 \end{eqnarray}
Since $a_{\mu}u^{\mu}=0$, the value $|a_4|$ is also the magnitude
of the 3-acceleration in the rest frame of the observer. From Eq.
(\ref{a}) we achieve the important conclusion that the origin of
the dS C-metric, $\rho=0$ (or $r=0$), is being accelerated with a
constant acceleration $|a_4|$ whose value is precisely given by
the parameter $A$ that appears in the dS C-metric. Moreover, at
radius $\rho=\ell$ [or $y=y_+$ defined in equation (\ref{F1})] the
acceleration is infinite which corresponds to the trajectory of a
null ray. Thus, observers held at $\rho=$const see this null ray
as an acceleration horizon and they will never see events beyond
this null ray. This acceleration horizon coincides with the dS
cosmological horizon and has a non-spherical shape. For the
benefit of comparison with the $A=0$ dS spacetime, we note that
when we set $A=0$, Eq. (\ref{a}) says that the origin, $\rho=0$,
has zero acceleration and at radius $\rho=\ell$ the acceleration
is again infinite but now this is due only to the presence of  the
usual dS cosmological horizon which has a spherical shape.

\vspace{0.1 cm} {\it The 5-Dimensional description}:

In order to improve and clarify the physical aspects of the dS
C-metric we turn now into the 5D representation of the solution.

The dS spacetime can be represented as the 4-hyperboloid,
\begin{eqnarray}
-(z^0)^2+(z^1)^2+(z^2)^2+(z^3)^2+(z^4)^2=\ell^2,
\label{hyperboloid}
 \end{eqnarray}
in the 5D Minkowski embedding spacetime,
\begin{eqnarray}
 d s^2 = -(dz^0)^2+(dz^1)^2+(dz^2)^2+(dz^3)^2+(dz^4)^2.
 \label{dS}
 \end{eqnarray}
Now, the massless uncharged dS C-metric is a dS spacetime in
disguise and therefore our next task is to understand how the dS
C-metric can be described in this 5D picture. To do this we first
recover the massless uncharged dS C-metric described by Eq.
(\ref{metric-int}) and apply to it the coordinate transformation
\cite{PodGrif2}
\begin{eqnarray}
  \hspace{-0.3cm} & & \hspace{-0.3cm}
  z^0=\gamma^{-1}\sqrt{\ell^2-\rho^2}\,\sinh(\tau/\ell)\:,
  \;\;\;\;\; z^2=\gamma^{-1} \rho \sin\theta \cos\phi \:,
  \nonumber \\
  \hspace{-0.3cm} & & \hspace{-0.3cm}
  z^1=\gamma^{-1}\sqrt{\ell^2-\rho^2}\,\cosh(\tau/\ell)\:,
  \;\;\;\;\;  z^3=\gamma^{-1} \rho \sin\theta \sin\phi \:,
  \nonumber \\
  \hspace{-0.3cm} & & \hspace{-0.3cm}
  z^4=\gamma^{-1}[\sqrt{1+\ell^2A^2} \,\rho \cos\theta
  +\ell^2A]\:,
\label{dS to dS-c}
  \end{eqnarray}
where $\gamma$ is defined in Eq. (\ref{gamma}). Transformations
(\ref{dS to dS-c}) define an embedding of the massless uncharged
dS C-metric into the 5D description of the dS spacetime since they
satisfy Eq. (\ref{hyperboloid}) and take directly Eq.
(\ref{metric-int}) into Eq. (\ref{dS}).

\begin{figure}[t]
\includegraphics*[height=2.3in]{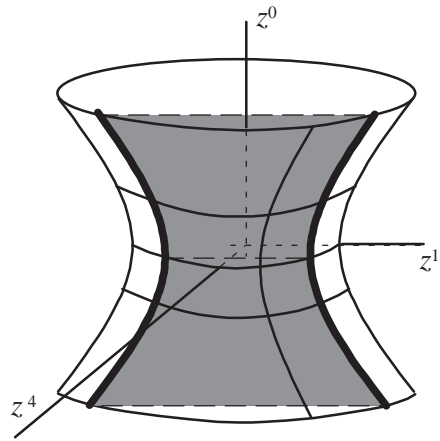}
\caption{\label{dS-hyperb} The dS 4-hyperboloid embedded in the 5D
Minkowski spacetime. The directions $z^2$ and $z^3$ are
suppressed. The two hyperbolic lines lying on the dS hyperboloid
result from the intersection of the hyperboloid surface with the
$z^4$=constant$<\ell$ plane. They describe the motion of the
origin of the dS C-metric ($A \neq 0$). For $A=0$ the intersecting
plane is $z^4=0$.
 }
\end{figure}

\begin{figure}[t]
\includegraphics*[height=2.2in]{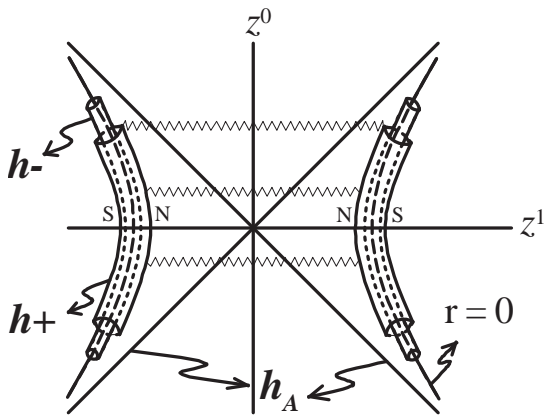}
\caption{\label{hyp} Schematic diagram representing the 5D
hyperbolic motion of two uniformly accelerating massive charged
black holes approaching asymptotically the Rindler-like
accelerated horizon ($h_A$). This horizon coincides with the
cosmological horizon. The inner and outer charged horizons are
represented by $h-$ and $h+$. The strut that connects the two
black holes is represented by the zigzag lines. The north pole
direction is represented by $\rm{N}$ and the south pole direction
by $\rm{S}$.
 }
\end{figure}

 So, the massless uncharged dS C-metric is a dS spacetime, but
we can extract more information from this 5D analysis. Indeed, let
us analyze with some detail the properties of the origin of the
radial coordinate, $\rho=0$ (or $r=0$). This origin moves in the
5D Minkowski embedding spacetime according to [see Eq. (\ref{dS to
dS-c})]
\begin{eqnarray}
 & & z^2=0\;,\;\; z^3=0\;,\;\; z^4=\ell^2A /
 \sqrt{1+\ell^2A^2}\:<\ell
 \;\;\;\;\mathrm{and} \nonumber \\
 & & (z^1)^2-(z^0)^2=(A^2+1/\ell^2)^{-1}\equiv a_5^{-2} \:.
\label{rindler}
  \end{eqnarray}
These equations define two hyperbolic lines lying on the dS
hyperboloid which result from the intersection of this hyperboloid
surface defined by Eq. (\ref{hyperboloid}) and the
$z^4$=constant$<\ell$ plane (see Fig. \ref{dS-hyperb}). They tell
us that the origin is subjected to a uniform 5D acceleration,
$a_5$, and consequently moves along a hyperbolic worldline in the
5D embedding space, describing a Rindler-like motion (see Figs.
\ref{dS-hyperb} and \ref{hyp}) that resembles the well-known
hyperbolic trajectory, $X^2-T^2=a^{-2}$, of an accelerated
observer in Minkowski space.
But uniformly accelerated radial worldlines in the 5D Minkowski
embedding space are also uniformly accelerated worldlines in the
4D dS space \cite{DesLev}, with the 5D acceleration $a_5$ being
related to the associated 4D acceleration $a_4$ by
$a_5^2=a_4^2+1/\ell^2$. Comparing this last relation with Eq.
(\ref{rindler}) we conclude that $a_4\equiv A$. Therefore, and
once again, we conclude that the origin of the dS C-metric is
uniformly accelerating with a 4D acceleration whose value is
precisely given by the parameter $A$ that appears in the dS
C-metric, Eq. (\ref{C-metric}), and this solution describes a dS
space whose origin is not at rest as usual but is being
accelerated. For the benefit of comparison with the $A=0$ dS
spacetime, note that the origin of the $A=0$ spacetime describes
the hyperbolic lines $(z^1)^2-(z^0)^2=\ell^2$ which result from
the intersection of the $z^4=0$ plane with the dS hyperboloid. In
this case we can say that we have two antipodal points on the
spatial 3-sphere of the dS space accelerating away from each other
due only to the cosmological background acceleration. When $A \neq
0$ these two points suffer an extra acceleration. This discussion
allowed us to find the physical interpretation of parameter $A$
and to justify its label. Notice also that the original dS
C-metric coordinates introduced in Eq. (\ref{C-metric}) cover only
the half-space $z^1>-z^0$. The Kruskal construction done in
section \ref{sec:PD} extended this solution to include also the
$z^1<-z^0$ region and so, in the extended solution, $r=0$ is
associated to two hyperbolas that represent two accelerated points
(see Fig. \ref{hyp}). These two hyperbolas approach asymptotically
the Rindler-like acceleration horizon ($r_A$).

\subsection{\label{sec:PI-mass} \textbf{Pair of accelerated black
holes ($\bm{m >0}$, $\bm{q \neq 0}$)}}
Now, we are in a position to interpret the massive and charged
solutions that describe two black holes accelerating away from
each other. To see clearly this, let us look to the Carter-Penrose
diagrams, Fig. \ref{Fig-1}, Fig. \ref{Fig-2} and Fig. \ref{Fig-3}.
Looking at these diagrams we can compare the different features
that belong to the massless uncharge case (Fig. \ref{Fig-1}), to
the non-extreme massive uncharged case [Fig. \ref{Fig-2}.(i)], and
ending in the non-extreme massive charged case [Fig.
\ref{Fig-3}).(i)]. In Fig. \ref{Fig-1} we identify the two
hyperbolas $r=0$ (represented by two timelike lines) approaching
asymptotically the Rindler-like acceleration horizon ($r_A$). When
we add a mass to the solution we conclude that each of these two
simple hyperbolas $r=0$ are replaced by the more complex structure
that represents a Schwarzschild black hole with its spacelike
curvature singularity and its horizon [this is represented by
$r_+$ in the left and right regions of Fig. \ref{Fig-2}.(i)]. So,
the two accelerating points $r=0$ have been replaced by two
Schwarzschild black holes that approach asymptotically the
Rindler-like acceleration horizon [represented by $r_A$ in the
middle region of Fig. \ref{Fig-2}.(i)]. The same interpretation
can be assigned to the massive charged solution. The two
hyperbolas $r=0$ of Fig. \ref{Fig-1} are replaced by two
Reissner-Nordstr\"{o}m black holes [with its timelike curvature
singularity and its inner $r_-$ and outer $r_+$ horizons; see the
left and right regions of Fig. \ref{Fig-3}.(i)] that approach
asymptotically the Rindler-like acceleration horizon already
present in the $m=0$ and $q=0$ causal diagram. An issue that is
relevant here, is whether the Cauchy horizons of the charged dS
C-metric are stable. The Cauchy horizon of the non-accelerated
dS$-$Reissner-Nordstr\"{o}m black holes
 is stable to small perturbations, as shown in \cite{stability} (for a review on
Cauchy horizon instabilities see, e.g., Burko and Ori
\cite{BurkoOri}). Moreover, in \cite{HorShein} it has been shown
that nearly extremal accelerating black holes in the flat
background have stable Cauchy horizons, unlike the Cauchy horizon
of the non-accelerated flat Reissner-Nordstr\"{o}m black hole
which is unstable. Therefore, we expect that the Cauchy horizons
of the accelerated dS$-$Reissner-Nordstr\"{o}m black holes are
stable, although we do not have confirmed this result. The
discussion of this subsection also applies directly to the extreme
cases of the dS C-metric.

\subsection{\label{sec:PI-strut} \textbf{Source of acceleration
and radiative properties}}
In this subsection we address the issue of the acceleration source
and its localization. In the massless uncharged dS C-metric,
observers that move along radial worldlines with $\rho=$const and
$\theta=0$ describe the Rindler-like hyperbola [see Eq. (\ref{dS
to dS-c})]
\begin{eqnarray}
(z^1)^2-(z^0)^2 &=& \frac{\ell^2-\rho^2}{(\sqrt{1+\ell^2A^2} +
A\rho)^2}  \:.
 \label{rindler.2}
  \end{eqnarray}
Moreover, when we put $m$ or $q$ different from zero, each of the
two hyperbolas assigned to $r=0$ represents the accelerated motion
of a black hole. Thus, from Eq. (\ref{rindler.2}) we conclude
\cite{OscLem_AdS-C} that the north pole axis is in the region
between the two black holes (see Fig. \ref{hyp}). Now, the value
of the arbitrary parameter $\kappa$ introduced in section
\ref{sec:Int} can be chosen in order to avoid a conical
singularity at the south pole ($\delta_\mathrm{s}=0$), leaving a
conical singularity at the north pole ($\delta_\mathrm{n}<0$).
This is associated to a strut that joins the two black holes along
their north poles and provides their acceleration
\cite{OscLem_AdS-C}. This strut satisfies the relation $p=-\mu>0$,
where $p$ and $\mu$ are respectively its pressure and its mass
density \cite{OscLem_AdS-C}. Alternatively, we can choose $\kappa$
such that avoids the deficit angle at the north pole
($\delta_\mathrm{n}=0$) and leaves a conical singularity at the
south pole ($\delta_\mathrm{s}>0$). This option leads to the
presence of a string (with $p=-\mu<0$) that connects the two black
holes along their south poles, and furnishes the acceleration.

\vspace{0.2 cm}

The C-metric is an exact solution that emits gravitational and
electromagnetic radiation. In the flat background the Bondi news
functions have been explicitly calculated in \cite{AshtDray,BPP}.
In dS background these calculations have not been carried yet, in
fact dS still lacks a peeling theorem.

\vspace{0.2 cm}

Ernst \cite{Ernst} has employed a Harrison-type transformation to
the $\Lambda=0$ charged C-metric in order to append a suitably
chosen external electromagnetic field. With this procedure the so
called Ernst solution is free of conical singularities at both
poles and the acceleration that drives away the two oppositely
charged Reissner-Nordstr\"{o}m black holes is totally provided by
the external electromagnetic field. In the dS background we cannot
remove the conical singularities through the application of the
Harrison transformation \cite{Emparan}. Indeed, the Harrison
transformation does not leave invariant the cosmological term in
the action. Therefore, applying the Harrison transformation to
Eqs. (\ref{C-metric})-(\ref{potential}) does not yield a new
solution of the Einstein-Maxwell-dS theory.

\section{\label{sec:Conc}SUMMARY AND CONCLUSIONS}
In a previous paper \cite{OscLem_AdS-C} we have analyzed in detail
the physical interpretation and properties of the anti-de Sitter
C-metric. In the dS background, Podolsk\'y and Griffiths
\cite{PodGrif2} have established that the dS C-metric found by
Pleba\'nski and Demia\'nski \cite{PlebDem} describes a pair of
accelerated black holes with the acceleration being provided by a
strut that connects the black holes together with the cosmological
constant. We have extended their analysis essentially by drawing
the Carter-Penrose diagrams of the solutions. These diagrams
allowed us to clearly identify the presence of two dS black holes
and to conclude that they cannot interact gravitationally. To
obtain the physical interpretation of the solutions we have
followed the approach of Kinnersly and Walker \cite{KW} for the
flat C-metric. The alternative approach of Bonnor \cite{Bonnor1}
which puts the flat C-metric into the Weyl form cannot be realized
here, since the introduction of the cosmological constant prevents
such a coordinate transformation.

The embedding of the massless uncharged dS C-metric into 5D
Minkowski space clearly shows that the origin of the dS C-metric
solution is subjected to a uniform acceleration, and describes a
hyperbolic Rindler-like worldline in the dS 4-hyperboloid embedded
in the 5D Minkowski space. To be more precise, the origin is
represented by two hyperbolic lines that approach asymptotically
the Rindler-like accelerated horizon. This acceleration horizon
coincides with the cosmological horizon of the dS solution that is
already present in the $A=0$ solution. So, the presence of the
extra acceleration caused by the strut does not introduce an extra
horizon, contrary to what occurs in the the $\Lambda<0$ C-metric
\cite{OscLem_AdS-C} and in the $\Lambda=0$ C-metric \cite{KW}.
However, in the $A=0$ dS solution the cosmological horizon has the
topology of a round sphere, while in the dS C-metric ($A\neq 0$)
the presence of the acceleration induces a non-spherical shape in
the acceleration (cosmological) horizon.

When we add a mass or a charge to the system the causal diagrams
indicate that now we have two dS-Schwarzschild or two
dS$-$Reissner-Nordstr\"{o}m black holes approaching asymptotically
the Rindler-like accelerated horizon. The topology of all the
horizons of these solutions is compact but non-spherical. In
\cite{PodGrif2} the authors stated that since the origin of the
massless uncharged dS C-metric corresponds to two accelerating
points in de dS background then, when the mass and charge are
small, the dS C-metric can be regarded as a perturbation of the
massless uncharged dS C-metric. In this way they concluded that
the dS C-metric describes a pair of accelerating black holes. With
the Carter-Penrose diagrams of the massive and charged solutions,
we have confirmed this result and we have found the range of
parameters $\Lambda, A, m$, and $q$ that correspond to non-extreme
black holes, extreme black holes, and naked particles. The general
features of the Carter-Penrose diagram of the dS C-metric are
independent of the angular coordinate $x$. This sets a great
difference between the causal diagrams of the dS C-metric and the
ones of the $\Lambda<0$ case \cite{OscLem_AdS-C} and of the
$\Lambda=0$ case \cite{KW}. Indeed, for the $\Lambda\leq 0$
C-metric, the Carter-Penrose diagram at the north pole direction
is substantially different from the one along the south pole
direction and different from the diagram along the equator
direction (see \cite{KW,OscLem_AdS-C}).

We have proceeded to the localization of the conical singularity
present in the solution. We have clearly identified the black hole
north pole direction and shown that it points towards the other
black hole. When the conical singularity is at the north pole, it
is associated to a strut between the two black holes which
satisfies the relation $p=-\mu>0$, where $p$ and $\mu$ are
respectively the pressure on the strut and its mass density. The
pressure is positive, so it points outwards and pulls the black
holes apart, furnishing their acceleration (as in the flat and AdS
C-metric). When the conical singularity is at the south pole, it
is associated to a string between the two black holes with
negative pressure that pushes the black holes away from each
other.

The analysis of the Carter-Penrose diagrams of the dS C-metric has
also been important to  conclude that the two black holes cannot
interact gravitationally. For example, looking into Fig.
\ref{Fig-2}.(i) which represents the non-extreme massive uncharged
C-metric we conclude that a null ray sent from the vicinity of one
of the black holes can never cross the acceleration horizon
($r_A$) into the other black hole. Indeed, recall that in these
diagrams the vertical axes represents the time flow and that light
rays move along $45^{\underline{\rm o}}$ lines. Thus, a light ray
that is emitted from a region next to the horizon $r_+$ of the
left black hole of Fig. \ref{Fig-2}.(i) can pass through the
acceleration horizon $r_A$ and, once it has done this, we will
necessarily proceed into infinity $\cal{I^+}$. This light ray
cannot enter the right region of Fig. \ref{Fig-2}.(i) and, in
particular, it cannot hit the horizon $r_+$ of the right black
hole. So, if the two black holes cannot communicate through a null
ray they cannot interact gravitationally. The black holes
accelerate away from each other only due to the pressure of the
strut that connects the two black holes, in addition to the
cosmological background acceleration contribution. This is
coherent with the fact that there is no particular value of
$\Lambda$ and $A$ for which the solution describes two black holes
in a dS background whose relative distance remains fixed. Indeed,
if the black holes could interact gravitationally then there would
exist a particular set of parameter values for which the outward
repulsion produced by the strut and by the cosmological constant
would exactly cancel the inward gravitational attraction, and the
two black holes would be in equilibrium in the dS background. This
is not the case precisely because the two black holes do not
interact gravitationally. Note that the dS-Schwarzschild solution
($A=0$) and the dS$-$Reissner-Nordstr\"{o}m solution ($A=0$) can
be interpreted as a pair of gravitationally non-interacting black
holes at antipodal points on the spatial 3-sphere of the dS space,
which are accelerating away from each other due to the
cosmological background acceleration. Similarly the massive and/or
charged dS C-metric ($A\neq 0$) also describes a pair of
gravitationally non-interacting black holes that accelerate away
from each other in a dS background, but now they suffer an extra
acceleration $A$ provided by the strut or by the string that
connects them.


\begin{acknowledgments}

This work was partially funded by Funda\c c\~ao para a Ci\^encia e
Tecnologia (FCT) through project CERN/FIS/43797/2001 and
PESO/PRO/2000/4014.  OJCD also acknowledges finantial support from
FCT through PRAXIS XXI programme. JPSL thanks Observat\'orio
Nacional do Rio de Janeiro for hospitality.

\end{acknowledgments}

\appendix*
\section{\label{sec:Phys_Interp m,e}Mass and charge parameters}
In this Appendix, one gives the physical interpretation of
parameters $m$ and $q$ that appear in the dS C-metric. We follow
\cite{PodGrif2}. Applying to  Eq. (\ref{C-metric}) the coordinate
transformations, $\tau=\sqrt{1+\ell^2A^2}A^{-1} t$,
$\rho=\sqrt{1+\ell^2A^2}(Ay)^{-1}$, together with  Eq. (\ref{ang})
and setting $A=0$ (and $\kappa=1$) one obtains
\begin{equation}
 d s^2 = - F(\rho)\, d \tau^2 +F^{-1}(\rho)\, d \rho^2
 +\rho^2 (d \theta^2+\sin^2\theta\,d\phi^2) \:,
 \label{mq2}
\end{equation}
where $F(\rho)=1-\rho^2/\ell^2 -2m/\rho + q^2/\rho^2$. So, when
the acceleration parameter vanishes, the dS C-metric, Eq.
(\ref{C-metric}), reduces to the dS-Schwarzschild and
dS$-$Reissner-Nordstr\"{o}m black holes and the parameters $m$ and
$q$ that are present in the dS C-metric are precisely the ADM mass
and ADM electromagnetic charge of these non-accelerated black
holes.


\end{document}